# A new metal transfer process for van der Waals contacts to vertical Schottky-junction transition metal dichalcogenide photovoltaics

Cora M. Went[1,2], Joeson Wong[3], Phillip R. Jahelka[3], Michael Kelzenberg[3], Souvik Biswas[3], Harry A. Atwater[2,3,4]*

1. Department of Physics, California Institute of Technology, Pasadena, CA 91125, USA
2. Resnick Sustainability Institute, California Institute of Technology, Pasadena, CA 91125, USA
3. Thomas J. Watson Laboratory of Applied Physics, California Institute of Technology, Pasadena, CA 91125, USA
4. Joint Center for Artificial Photosynthesis, California Institute of Technology, Pasadena, CA 91125, USA

*Corresponding author: Harry A. Atwater (haa@caltech.edu)

## Abstract

Two-dimensional transition metal dichalcogenides are promising candidates for ultrathin optoelectronic devices due to their high absorption coefficients and intrinsically passivated surfaces. To maintain these near-perfect surfaces, recent research has focused on fabricating contacts that limit Fermi-level pinning at the metal-semiconductor interface. Here, we develop a new, simple procedure for transferring metal contacts that does not require aligned lithography. Using this technique, we fabricate vertical Schottky-junction $WS_2$ solar cells with Ag and Au as asymmetric work function contacts. Under laser illumination, we observe rectifying behavior and open-circuit voltage above 500 mV in devices with transferred contacts, in contrast to resistive behavior and open-circuit voltage below 15 mV in devices with evaporated contacts. One-sun measurements and device simulation results indicate that this metal transfer process could enable high-specific-power vertical Schottky-junction transition metal dichalcogenide photovoltaics, and we anticipate that this technique will lead to advances for two-dimensional devices more broadly.

Two-dimensional (2D) semiconducting transition metal dichalcogenides (TMDs), including MoS$_2$, WS$_2$, MoSe$_2$, and WSe$_2$, are promising for many optoelectronic applications, including high-specific-power photovoltaics[1–4]. With absorption coefficients 1–2 orders of magnitude higher than conventional semiconductors, monolayer (<1 nm thick) TMDs can absorb as much visible light as about 15 nm of GaAs or 50 nm of Si[5]. Both multilayer and monolayer TMDs can achieve near-unity broadband absorption in the visible range[6,7]. Due to their layered structure and out-of-plane van der Waals bonding, TMDs have intrinsically passivated surfaces with no dangling bonds and can form heterostructures without the constraint of lattice matching.

To take advantage of the intrinsically passivated surfaces of TMDs, gentle fabrication techniques are needed to form metal contacts without damaging the underlying semiconductor. A number of new contact techniques have been presented recently, including one-dimensional edge contacts[8], via contacts embedded in hBN[9], slowly-deposited In/Au contacts[10], and 2D metals[11]. Recently, Liu *et al* have shown that transferring rather than evaporating metal contacts onto TMDs can yield interfaces with no Fermi-level pinning, where the Schottky barrier height can be predicted by the ideal Schottky-Mott rule[12]. Their work demonstrates the utility of transferring an arbitrary three-dimensional metal onto a 2D material, forming a nondamaging van der Waals contact. However, this technique requires a final aligned lithography step to expose the contact under the polymer used for transfer, which limits its scalability[12].

To date, the above techniques for gentle contact fabrication have been applied to device geometries where carriers are collected laterally rather than vertically. Though laterally-contacted devices are important for electronic applications, such as field-effect transistors, vertically-contacted devices are necessary for optoelectronic applications that require scalable photoactive areas, such as solar cells. Van der Waals contacts could have an even greater advantage for these vertical device geometries, where the ratio of contact area to device area is often higher than in lateral device geometries.

Schottky-junction solar cells represent one specific device geometry where van der Waals metal contacts could enable high performance in vertical devices. Although vertical-junction solar cells are more aligned with conventional photovoltaics[13], most Schottky-junction TMD solar cells studied have been lateral-junction devices[12,14,15]. Vertical Schottky-junction TMD solar cells have been limited by ohmic I-V behavior, low external quantum efficiencies, and low open-circuit voltages, likely due to Fermi-level pinning induced by contact evaporation[6,16]. New gentle contact fabrication techniques have the potential to eliminate this Fermi-level pinning, enabling high-efficiency vertical TMD solar cells in the Schottky-junction geometry.

Here, we develop a simple technique for transferring metal contacts, where all lithographic patterning is done on a donor substrate rather than on the active device. We apply this technique to vertical Schottky-junction solar cells with multilayer TMD absorber layers. Due to the tradeoff between bandgap energy and photoluminescence quantum yield, the theoretical maximum power conversion efficiency achievable for multilayer and monolayer single-junction solar cells is similar[4,17], and further, tunneling limits transport in monolayer vertical devices[18], so we focus on multilayer devices in this work. Ultrathin (10–20 nm) $WS_2$ forms the absorber layer, while Ag and Au form the asymmetric-work-function contacts. Devices made with transferred metal contacts show diode-like I-V behavior with a near-unity ideality factor and high $V_{OC}$, while similar devices made with evaporated metal contacts show ohmic I-V behavior and near-zero $V_{OC}$. We demonstrate peak external quantum efficiency (EQE) of >40% and peak active-layer internal quantum efficiency ($IQE_{active}$) of >90% in transferred-contact devices. Using a solar simulator, we measure a photovoltaic power conversion efficiency of 0.46%, comparable to what has been seen in other ultrathin vertical TMD photovoltaics[19]. Device simulations of further-optimized geometries suggest that this new metal transfer process has the potential to enable Schottky-junction TMD solar cells with power conversion efficiencies greater than 8% and specific powers greater than 50 kW/kg.

**Results & Discussion**

**Fabrication of vertical $WS_2$ Schottky-junction solar cells.** We prepare vertical $WS_2$ Schottky-junction solar cells made from 16-nm-thick $WS_2$ absorber layers, with Ag ($\phi_{Ag} \approx 4.3$ eV) and Au ($\phi_{Au} \approx 5.1$ eV) as asymmetric work function contacts (Fig. 1a)[12]. Template-stripped Ag, which exhibits an RMS roughness <0.5 nm, forms both the electron-collecting bottom contact and back reflector for all devices[20]. We mechanically exfoliate $WS_2$ directly onto the Ag substrate. The subwavelength-thick $WS_2$ achieves broadband, angle-insensitive absorption on top of the highly reflective Ag, giving the $WS_2$ a deep purple color as illustrated in Fig. 1b[6,21]. For transferred-contact devices, we transfer thin Au disks from a thermally-oxidized Si donor substrate to form the semi-transparent hole-collecting top contact, using the process described in the following section. Both the top surface of the template-stripped Ag and the bottom surface of the transferred

Au inherit the smoothness of the SiO$_2$/Si donor substrate, leading to near atomically-sharp metal-WS$_2$ interfaces[12,20]. For comparison, we also fabricate devices by direct evaporation of thin Au disks onto the WS$_2$ using standard photolithography techniques.

The ideal band diagram of this Schottky-junction solar cell is shown in Fig. 1c. We assume a doping concentration of $10^{14}$ cm$^{-3}$ for WS$_2$, as provided by the bulk crystal vendor. Since the length of the depletion region at a Schottky junction between bulk WS$_2$ and either Au or Ag is on the order of 1 μm, the device is fully depleted. We measure the final thicknesses of the WS$_2$ and the Au to be 16 nm and 19 nm, respectively, using atomic force microscopy (Fig. 1d).

**Metal transfer process.** We develop a new, simple process for transferring metal contacts onto TMDs (Fig. 2). This process relies on a self-assembled monolayer (SAM) to reduce the adhesion between the Au and the SiO$_2$/Si donor substrate[22], a thermoplastic polymer to preferentially pick up or drop down the metal[23], and a variable peeling rate to tune the velocity-dependent adhesion between a metal and a viscoelastic stamp[22].

Briefly, we create a SAM on clean thermally-oxidized Si chips in a vacuum desiccator[22]. We then deposit 20 nm of Au via electron beam evaporation. Using photolithography, we define the contact areas with positive photoresist and a positive photomask. We etch the Au outside the masked contact areas, then dissolve the remaining photoresist in acetone, leaving Au disks on the SAM-coated SiO$_2$/Si substrates. We prepare a polydimethylsiloxane (PDMS) stamp coated with the thermoplastic polymer polypropylene carbonate (PPC) on a glass slide[23]. In a 2D transfer setup, we align and slowly lower the stamp onto a contact at 60ºC. We set the temperature to 40ºC, and once the stage reaches that temperature, we raise the transfer arm rapidly to peel the stamp and pick up the contact. We align the contact with the target TMD and slowly lower the stamp down at 60ºC, and then slowly peel it away immediately after contact at the same temperature. The contact delaminates from the PDMS/PPC stamp and sticks to the TMD. Further details of the procedure are provided in Supplementary Note 1.

This metal transfer technique has worked in 15 out of 16 devices fabricated thus far (94% yield). It works for both 20-nm-thick and 100-nm-thick Au, and can likely be extended to other metals and to larger-scale contacts (i.e. for contacts to CVD-grown TMDs). A substantial advantage of this technique is that, whereas prior metal transfer techniques require a final aligned electron-beam lithography step to expose the contact area[12], this technique only utilizes unaligned photolithography to define the initial contacts on the SiO$_2$/Si donor substrate. This allows for batch-fabrication of an array of contacts that can then be picked up, aligned, and printed to form multiple devices. Further, this metal transfer process could enable van der Waals contacts to air- and moisture-sensitive nanomaterials, such as lead halide perovskites or black phosphorus, to be formed without removing the sample from an inert environment.

**Comparison of transferred and evaporated metal contacts.** We measure I-V curves under illumination with a 633 nm laser focused to a ~1 μm$^2$ spot in a confocal microscope at room temperature. In devices with transferred metal contacts, we observe rectifying I-V curves and a pronounced photovoltaic effect (Fig. 3a). We measure a $V_{OC}$ of 510 mV under the maximum laser excitation. Short-circuit current follows a power law as a function of incident power, $I_{SC} = P_{inc}^\alpha$, with α close to 1 (Fig. 3c). According to the diode equation, $V_{OC}$ scales linearly with ln($I_{SC}$) and can be fit with an ideality factor n = 1.2 (Fig. 3d). This near-unity ideality factor confirms the high interface and material quality in these devices. The ideality factor, diode-like behavior and high open-circuit voltage suggest that a Schottky junction is successfully formed in devices with transferred contacts.

In contrast, we observe resistive behavior and a small photovoltaic effect in devices with evaporated top metal contacts (Fig. 3b). $I_{SC}$ vs. $P_{inc}$ follows a power law with α less than 1 (Fig. 3e). As shown in Fig. 3f, this device behaves as a resistor with R = 3.1 kΩ. At comparable laser powers, $V_{OC}$ is around 4 mV in evaporated contact devices and 400 mV in transferred contact devices, and $J_{SC}$ is three to four times higher for transferred contacts than for evaporated contacts. Previous work demonstrates that due to Fermi-level pinning, evaporated Au and transferred Ag have effectively the same barrier height for electrons and holes[12]. Assuming an effective work function difference between Au and Ag of 50 meV, device simulations can predict the purely resistive behavior in an evaporated-contact Schottky-junction device (Supplementary Fig. 1). This evidence points to strong Fermi-level pinning in devices with evaporated contacts due to interface states induced by the Au evaporation.

In devices with transferred contacts, the slope of the I-V curve at short-circuit increases linearly with increasing laser power, corresponding to a decreasing shunt resistance (Supplementary Fig. 2a). This photoshunting effect occurs in solar cells without perfectly selective contacts due to increased minority carrier conductivity across the device under illumination[24,25].

Device simulations can replicate this photoshunt pathway without the addition of any external shunt resistance (Supplementary Fig. 2b). In future devices, the introduction of contacts with greater carrier selectivity could reduce or eliminate the photoshunting observed here.

**Quantum efficiency and photocurrent generation.** Light-beam induced current (LBIC, or photocurrent) maps, acquired with a 633 nm laser in a confocal microscope, show uniform current generation under the entire Au disk contact, except where shaded by the contact probe (Fig. 4a). The uniformity of the photocurrent demonstrates that the Au is homogeneously semitransparent and in good contact with the TMD. Importantly, this indicates that the area of the Au disk can be used to accurately define the device active area (Supplementary Fig. 3) and suggests that one-dimensional device simulations are sufficient to describe the behavior in these vertical devices[26]. Further, it demonstrates that there are no visible bubbles created during the metal transfer process.

The measured total absorption (Fig. 4c) matches well with the absorption calculated using the transfer matrix method (Fig. 4b), as has been previously demonstrated in TMD solar cells[6,16]. To calculate the active-layer absorption in the experimental $WS_2$ devices, we subtract the simulated parasitic absorption (the sum of the Au and Ag curves in Fig. 4b) from the experimentally measured total absorption in Fig. 4c. The mean active-layer absorption from 450 nm to 650 nm is 39%. The reduced absorption in our devices relative to what has been previously demonstrated in $WS_2$ on a metal back-reflector (~80% over this wavelength range)[6] is due to parasitic absorption and reflection losses from the 19-nm-thick Au top contact. Using a more transparent top contact could double our photogenerated current, assuming identical work function and conductivity.

The external quantum efficiency (EQE) of the device follows the spectral shape of absorption well, averaging 28% from 450 nm to 650 nm and reaching a peak of above 40% around 550 nm (Fig. 4d). To accurately determine EQE, we multiply by a shading factor of 1.39 to correct for shading from the probes (see Methods). Internal quantum efficiency (IQE) remains relatively flat across all wavelengths above the bandgap, averaging 49% from 450 nm to 650 nm (Fig. 4e). $IQE_{active}$, calculated by dividing EQE by the active-layer absorption rather than the total absorption, is greater than 90% at its peak, and averages 74% between 450 nm and 650 nm (Fig. 4f). This high $IQE_{active}$ suggests efficient collection of photogenerated carriers in transferred-contact devices.

**Performance under one-sun illumination.** Vertical Schottky-junction WS$_2$ solar cells with transferred top contacts achieve reasonable photovoltaic performance when measured under simulated AM1.5G illumination. Fig. 5 shows the AM1.5G I-V behavior of a representative device. We divide the measured current by the device active area to yield current density, then then further divide by a factor of 0.67 to account for spectral mismatch between our solar simulator calibration point and the true AM1.5G spectrum (see Methods; Supplementary Fig. 4)[27]. The spectral mismatch correction leads to a 50% increase in short-circuit current, so the V$_{OC}$ and power conversion efficiency of the device are likely underestimated here. We measure a V$_{OC}$ of 256 mV, a corrected J$_{SC}$ of 4.10 mA/cm$^2$, a fill factor (FF) of 0.44, and a power conversion efficiency (PCE) of 0.46%. This efficiency is in the range of what others have reported for ultrathin TMD photovoltaics[15,16,19]. Using the densities of Au, WS$_2$, and Ag, we estimate a specific power of 3 kW/kg for this device.

By fitting the one-sun I-V curve using the diode equation with series and shunt resistances, we estimate a shunt resistance (R$_{SH}$) of 231 Ω cm$^2$ and a negligible series resistance (R$_S$), as shown in Supplementary Fig. 5. The shunt resistance is likely due to the photoshunting behavior discussed above, and could be reduced by design and realization of contacts that are more carrier-selective.

This photovoltaic performance is consistent among multiple measurements and devices. The J$_{SC}$ of 4.10 mA/cm$^2$ that we measure with the solar simulator is within 10% of the J$_{SC}$ that we calculate by integrating the EQE over the solar spectrum (4.55 mA/cm$^2$). We believe that probe shading, which we correct for in EQE measurements but not in solar simulator measurements, accounts for the 10% discrepancy. Though the J$_{SC}$ varies due to differences in thickness and therefore absorption in exfoliated flakes, the V$_{OC}$ is replicable across all devices fabricated for this work. As shown in Supplementary Figs. 6 and 7, V$_{OC}$ is between 220 mV and 260 mV in all four devices measured under one-sun illumination, and V$_{OC}$ is greater than 220 mV in six different devices measured under illumination with a halogen lamp (~20 suns power density). The I-V curves show no hysteresis when swept in the forwards and backwards directions (Supplementary Fig. 8).

**Simulated performance of optimized devices.** To examine and further optimize the performance of these devices, we simulate a variety of device geometries. The assumed material parameters of the WS$_2$ are detailed in Supplementary Table 1. Simulating the same device geometry as our experimental device yields the I-V curve in Fig. 6a. The simulated $J_{SC}$ of 5.7 mA/cm$^2$ is consistent with our measured active-layer IQE of 74% and the $J_{SC}$ of 4.55 mA/cm$^2$ estimated from the EQE. The simulated $V_{OC}$ of 646 mV and $R_{SH}$ of 2240 Ω cm$^2$ are considerably higher than the $V_{OC}$ of 256 mV and $R_{SH}$ of 231 Ω cm$^2$ observed in our one-sun measurements. This demonstrates that with further optimization, our device geometry could achieve higher voltages and less shunting than we currently see (Supplementary Fig. 9). As a first improvement, we suggest replacing Au with a different high-work-function metal, as Au is known to form thiol bonds with sulfides that could affect the quality of the van der Waals contact[12,28].

To identify a potential path towards high-efficiency vertical Schottky-junction WS$_2$ solar cells, we simulate a series of optimized devices (Fig. 6b). Using an optimized thickness of WS$_2$ (26 nm) for maximum absorption under 20 nm of Au increases the $J_{SC}$ to 7.1 mA/cm$^2$. The $J_{SC}$ can be further increased to 12.5 mA/cm$^2$ by replacing Au with a transparent top contact, assuming an identical work function to Au and no parasitic absorption or reflection. By selecting metal work functions that are optimally aligned to the conduction and valence bands of WS$_2$, we predict a $V_{OC}$ increase of 230 mV. Combining transparent top contacts and optimized metal work functions yields the device shown in Fig. 6c, with a $V_{OC}$ of 898 mV, a $J_{SC}$ of 12.7 mA/cm$^2$, a fill factor of 0.78, and a power conversion efficiency of 8.9%. This simulated power conversion efficiency in a device with a thickness <150 nm represents a specific power of 58 kW/kg, demonstrating that this metal transfer process has the potential to enable devices with an unprecedented power-per-unit-weight ratio for transportation and aerospace applications.

**Conclusion**

We develop here a process for transferring metal contacts with near-atomically smooth interfaces that has high-yield, allows for batch fabrication, and eliminates aligned lithography. We expect that this procedure will be highly relevant and useful to the 2D community, as well as to researchers working on air-sensitive nanomaterials, as it allows all processing to be done on the contacts rather than the device. By applying this new technique to vertical Schottky-junction TMD solar cells, we demonstrate that transferred contacts are particularly advantageous for vertical device geometries, which are important for photovoltaic and other optoelectronic applications due to their scalable active areas.

The rectifying I-V curves shown in transferred-contact devices and resistive I-V curves shown in evaporated-contact devices support the hypothesis that transferring contacts can reduce Fermi-level pinning and allow the work function asymmetry between the contacts to define the maximum achievable $V_{OC}$. We observe active-layer absorption >55%, EQE >40%, and active-layer IQE >90% in these devices, demonstrating efficient collection of photogenerated carriers. Under one-sun illumination, we measure a $V_{OC}$ of 256 mV, a $J_{SC}$ of 4.10 mA/cm$^2$, a fill factor of 0.44, and a power conversion efficiency of 0.46%. We highlight areas for improvement by simulating the behavior of optimized devices based on this architecture and show 8.9% simulated efficiency and 58 kW/kg simulated specific power in a device with transparent top contacts, optimized thickness, and ideal metal work functions for carrier extraction.

Given the proof-of-concept performance and the clear pathways for improvement presented here for devices less than 150 nm thick, ultrathin vertical Schottky-junction TMD solar cells with transferred contacts are promising candidates for high-specific-power photovoltaic applications. We anticipate that this new metal transfer process will enable similar advances for 2D TMD devices beyond Schottky-junction solar cells, as well as for nanomaterial-based devices more broadly.

**Methods**

**Device fabrication.** Template-stripped silver substrates are prepared as described previously[16,20]. WS$_2$ is mechanically exfoliated directly onto template-stripped silver from the bulk crystal (HQ Graphene) using Scotch tape. For transferred contact devices, Au top contacts are prepared and transferred using the metal transfer technique summarized in the main text, and described in detail in Supplementary Note 1. The SAM used is trichloro(1H,1H,2H,2H-perfluorooctyl)silane (PFOTS, Sigma Aldrich), the photoresist used is S1813, and the Au etchant used is Transene Gold Etchant TFA. For evaporated contact devices, Au top contacts are patterned using standard photolithography techniques as described previously[16]. Contacts are fabricated on WS$_2$ within 12 hours of exfoliation. Final WS$_2$ and Au thicknesses are confirmed using atomic force microscopy (Asylum Research).

**Photocurrent and power-dependent I-V.** Photocurrent and power-dependent IV are measured on a scanning confocal microscope (Zeiss Axio Imager 2) using a long working distance objective (50x, NA = 0.55). Devices are contacted using piezoelectrically controlled micromanipulators (MiBots, Imina Technologies). I-V curves are measured with a Keithley 236 Source-Measure Unit using custom LabView programs. Laser powers are measured using a USB power meter (ThorLabs). All measurements are performed under ambient temperature and pressure.

**Absorption and EQE.** Absorption and EQE are measured using a home-built optical setup with a long working distance objective (50x, NA = 0.55). A supercontinuum laser (Fianium) is coupled to a monochromator to produce a tunable, monochromatic light source. A chopper and lock-in detection are used for all measurements. For absorption, the sample reflectance is measured using a NIST-calibrated photodetector (Newport 818-ST2-UV/DB) with a beamsplitter. A protected silver mirror (Thorlabs) is used to calibrate the reflectance based on its reported reflectance curve, and a dark background is subtracted from both measurements. For EQE, the current generated by the sample is probed using MiBots and compared to the current collected by the NIST calibrated photodetector when placed at the sample position, corrected by the photodetector's responsivity. Absorption and EQE measurements are both corrected by a shading factor of 1.39 that corrects for the shading of the MiBot tips, which is calculated by comparing absorption with and without the tips in place and averaging over the spectral range 450 nm–650 nm.

**Solar simulator.** One-sun I-V curves are measured using a 1 kW Xenon arc lamp (Newport Oriel) with an AM1.5G filter (ABET Technologies). To ensure 100 mW/cm$^2$ incident power, the lamp power is adjusted to generate the correct current on a Si reference cell placed at the same location as the sample. MiBots are used to contact the device, and I-V curves are measured with a Keithley 2425 SourceMeter using custom LabView programs. The current density is divided by a spectral mismatch factor to account for the difference in bandgap between our WS$_2$ sample and our Si reference cell and the difference in spectrum between our solar simulator and AM1.5G[27]. As no EQE data was available for our device below 400 nm, linear extrapolation was used, leading to about a 5% error in the spectral mismatch factor and the reported J$_{SC}$ values. The device area was assumed to be that of the Au disk, which has a diameter of 28 µm.

**Device simulations.** Absorption and generation are calculated using the transfer matrix method, with optical constants for $WS_2$ taken from literature[29]. All other device simulations are performed using Lumerical Device. The $WS_2$ doping was specified by the bulk crystal vendor (HQ Graphene). Other $WS_2$ parameters, including bandgap[30], work function[31], DC permittivity[32], effective mass[33], out-of-plane mobility[34–36], and photoluminescence quantum yield[37–39] are taken from literature and listed in Supplementary Table 1. The radiative recombination coefficient is calculated using the Roosbroeck-Shockley relation[40], and the Shockley-Read-Hall lifetime for minority carriers is then estimated using the photoluminescence quantum yield.

**Data availability**

All data supporting the findings of this study are available from the corresponding author upon request.

**Acknowledgements**

This work was supported by the DOE 'Photonics at Thermodynamic Limits' Energy Frontier Research Center under grant DE-SC0019140. C.M.W. and J.W. acknowledge support from the National Science Foundation Graduate Research Fellowship under grants 1745301 and 1144469. C.M.W. acknowledges fellowship support from the Resnick Institute. The authors thank Sungwoo Nam for useful discussions.

**Author contributions**

C.M.W. fabricated the devices, performed the measurements, and performed the simulations. C.M.W., J.W., P.R.J., and S.B. developed the metal transfer technique. J.W. and P.R.J. assisted with the simulations. M.K. assisted with the solar simulator, absorption, and EQE measurements. H.A.A. supervised all experiments, calculations, and data collection. All authors contributed to data interpretation, presentation, and writing of the manuscript.

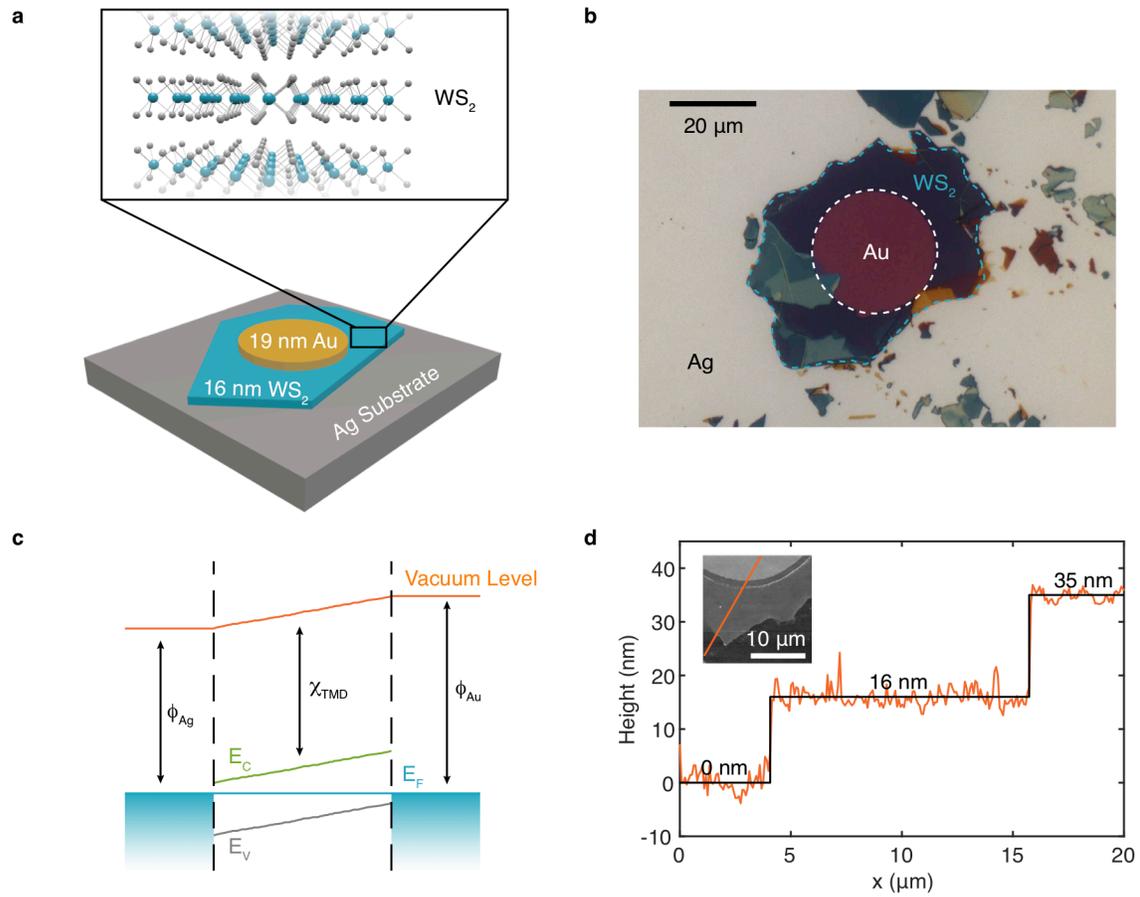

**Fig. 1** Vertical Schottky-junction multilayer WS$_2$ solar cells with transferred contacts. **a** Bottom: Schematic of device structure. Top contact is a transferred gold disk; bottom contact and back-reflector is template-stripped silver. Top: three-dimensional representation of multilayer WS$_2$. W, blue spheres; S, grey spheres. **b** Optical image of device. **c** Solar cell band diagram obtained from electrostatic simulations. **d** AFM height profile across device.

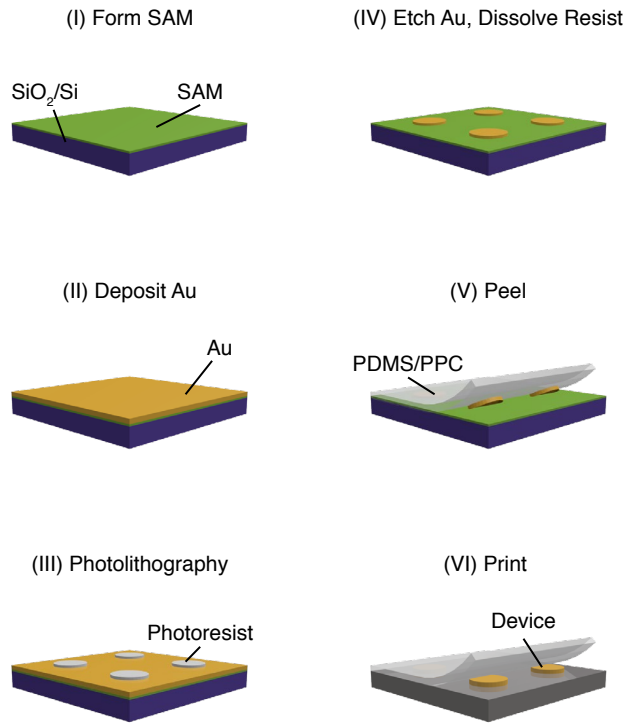

**Fig. 2** Metal transfer process. Briefly, a self-assembled monolayer is applied to a clean SiO$_2$/Si substrate (I). Gold is deposited in an electron beam evaporator (II). Disk contacts are defined using photolithography (III), and the surrounding gold is etched away (IV). To peel the contacts, a PDMS/PPC stamp is laminated to the contacts, heated above the glass transition temperature of PPC, then cooled and removed quickly (V). To print the contacts, the PDMS/PPC stamp is aligned and laminated onto the device, then peeled away slowly above the glass transition temperature of PPC, leaving the contacts behind (VI).

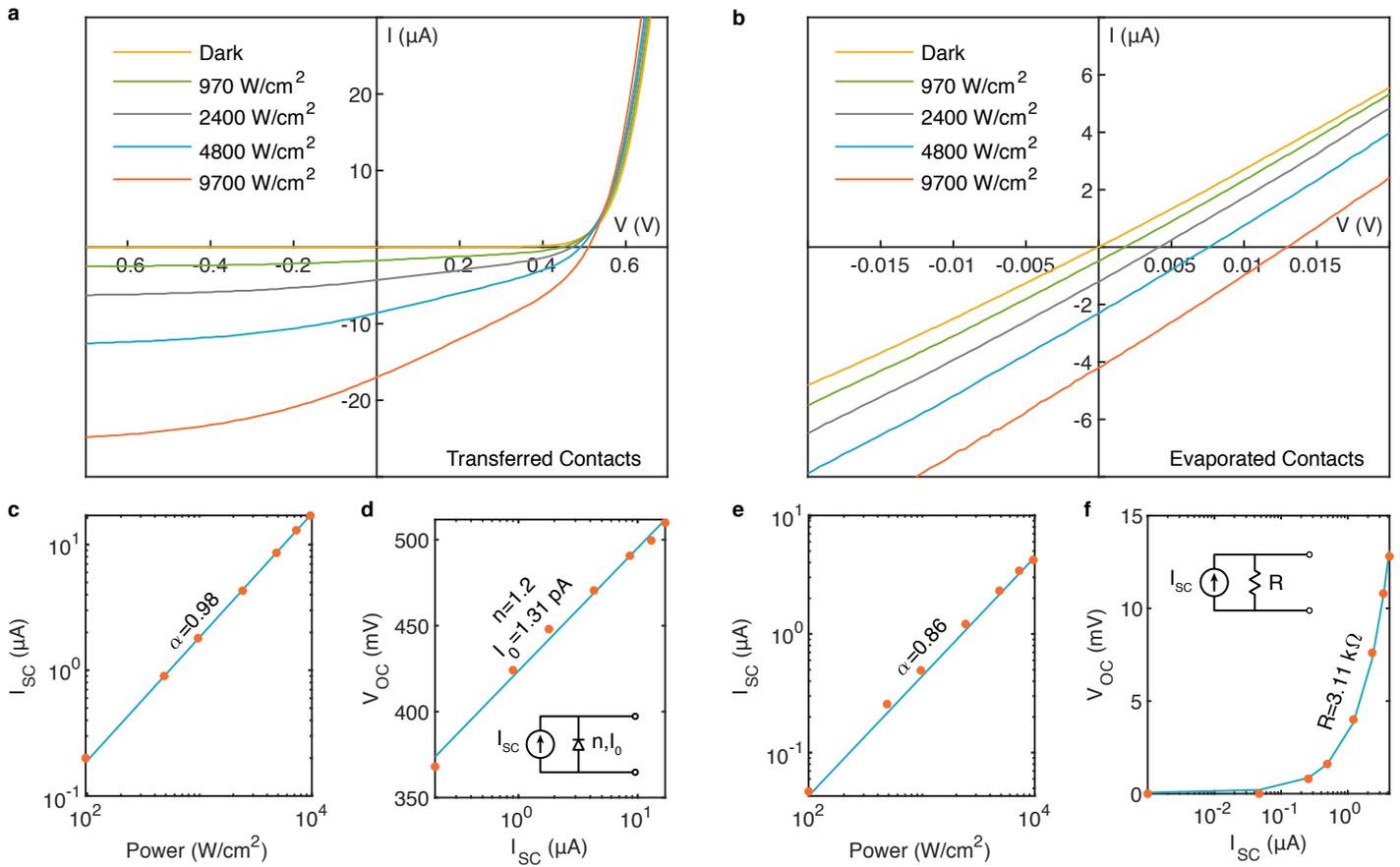

**Fig. 3** Comparison of devices with transferred and directly evaporated top metal contacts. **a, b** Power-dependent I-V characteristics of devices with transferred (**a**) and evaporated (**b**) gold top contacts taken under excitation with a 633 nm laser focused to a spot size of ~1 μm². **c, e** Short-circuit current of devices with transferred (**c**) and evaporated (**e**) gold contacts. Symbols, measurements; line, power law fit. **d, f** Open-circuit voltage of devices with transferred (**d**) and evaporated (**f**) gold contacts. Symbols, measurements; line, fit. Insets show representative circuit diagrams. n is the ideality factor and $I_0$ is the dark saturation current extracted from the diode fit in (**d**). R is the resistance extracted from the linear fit in (**f**).

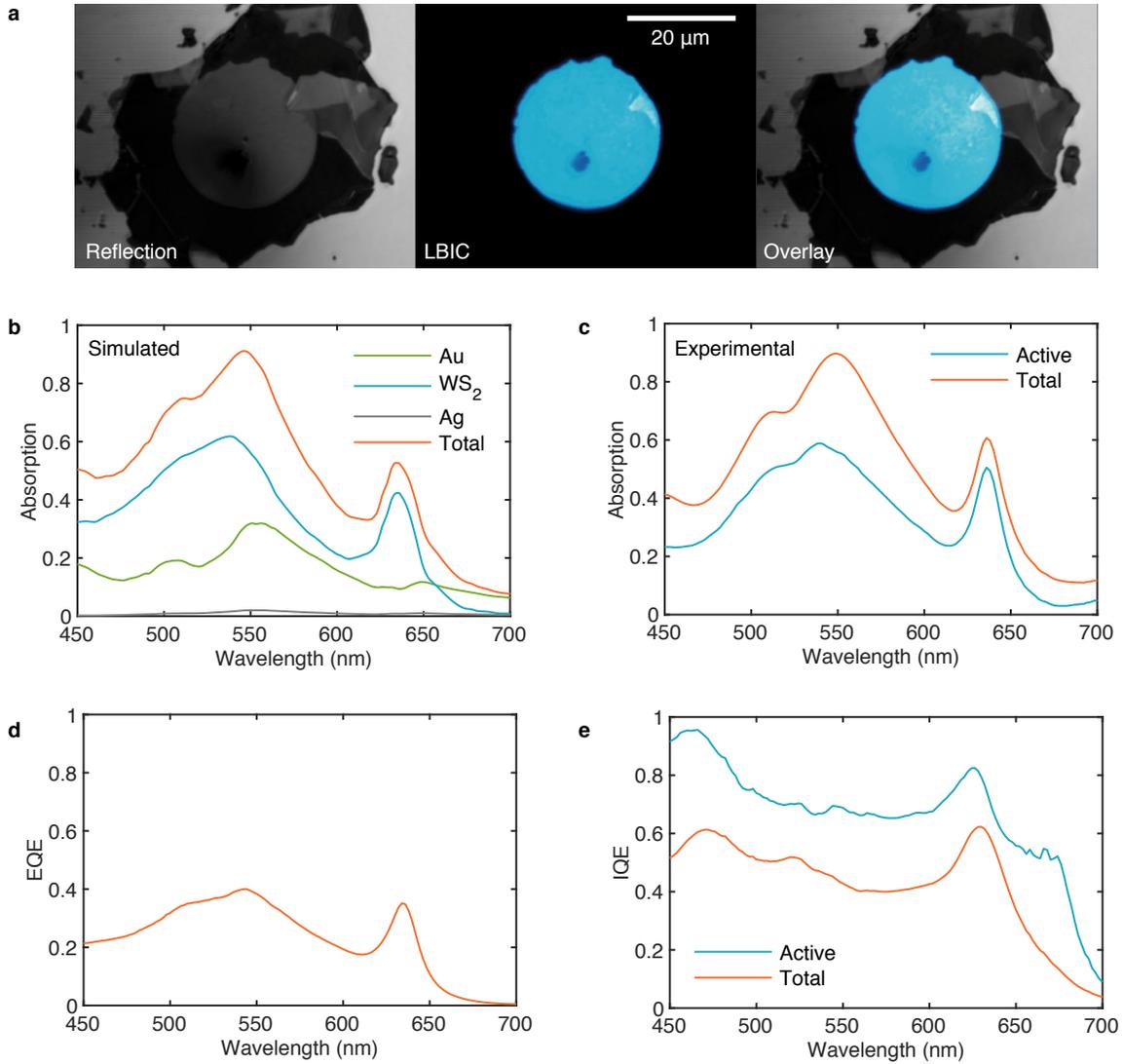

**Fig. 4** Photocurrent and quantum efficiency. **a** Confocal reflection, photocurrent, and reflection/photocurrent overlay maps for the device shown in Fig. 1. The dark spot in the lower left part of the device active area is a probe tip artifact. **b** Simulated total absorption in the device and absorption in each device layer. **c** Experimental total absorption in the device and active-layer absorption calculated by subtracting the simulated parasitic absorption from the experimentally measured total absorption. **d** Measured external quantum efficiency. **e** Internal quantum efficiency calculated from external quantum efficiency and absorption; active-layer internal quantum efficiency calculated from external quantum efficiency and active-layer absorption.

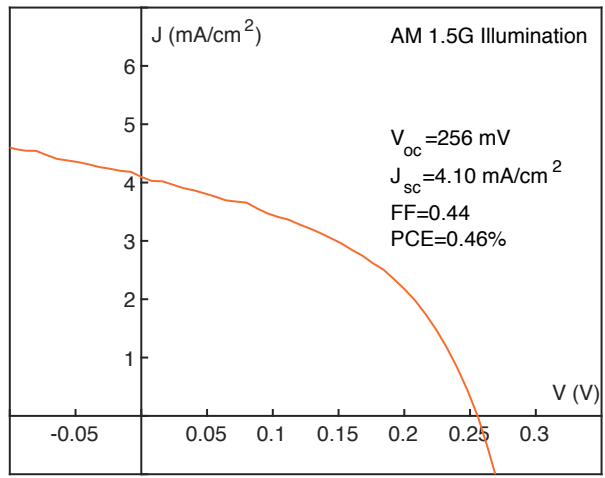

**Fig. 5** Photovoltaic performance under one-sun illumination. I-V characteristics of a vertical Schottky-junction multilayer $WS_2$ solar cell measured using an AM1.5G solar simulator, corrected for spectral mismatch.

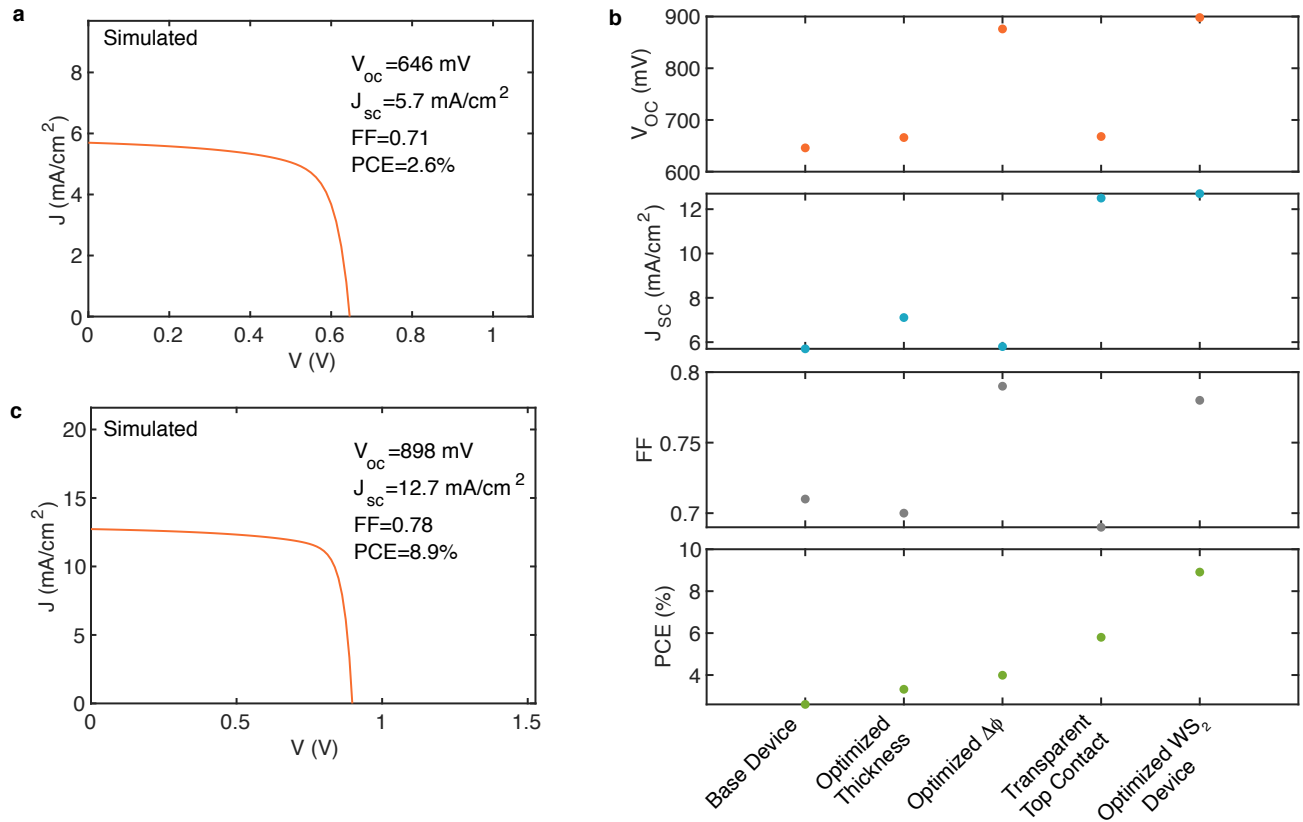

**Fig. 6** Simulated performance of optimized devices. **a** Simulated I-V characteristics of the device geometry used in our experiments, assuming no external series/shunt resistances. **b** Simulated $V_{OC}$, $J_{SC}$, fill factor, and power conversion efficiency for optimized devices. Apart from the final device geometry ("Optimized WS$_2$ Device"), optimizations are independent, not cumulative. Base device, the device geometry simulated in (**a**). Optimized thickness, the WS$_2$ thickness that maximizes $J_{SC}$ under 20 nm of Au (26 nm). Optimized $\Delta\phi$, metal work functions matched ideally to the WS$_2$ conduction and valence band edges ($\phi_1 = 4.05$ and $\phi_2 = 5.2$ eV). Transparent top contact, a top contact with no parasitic absorption/reflection and the same work function as Au on 16 nm of WS$_2$. Optimized WS$_2$ device, a device with optimized metal work functions, optimized WS$_2$ thickness, and transparent top contacts. **c** Simulated I-V characteristics of this optimized WS$_2$ device.

# Supplementary Information for "Transferred metal contacts for vertical Schottky-junction transition metal dichalcogenide photovoltaics"

Cora M. Went[1,2], Joeson Wong[3], Phillip R. Jahelka[3], Michael Kelzenberg[3], Souvik Biswas[3], Harry A. Atwater[2,3,4]*

1. Department of Physics, California Institute of Technology, Pasadena, CA 91125, USA
2. Resnick Sustainability Institute, California Institute of Technology, Pasadena, CA 91125, USA
3. Thomas J. Watson Laboratory of Applied Physics, California Institute of Technology, Pasadena, CA 91125, USA
4. Joint Center for Artificial Photosynthesis, California Institute of Technology, Pasadena, CA 91125, USA

*Corresponding author: Harry A. Atwater (haa@caltech.edu)

**Supplementary Note 1: Detailed Metal Transfer Procedure**

Complete details of the metal transfer procedure are described below.

*Step 1: Metal contacts on SAM-coated Si/SiO₂.* Commercially-available Si wafers coated with 285 nm thermal $SiO_2$ are diced into chips, cleaned in Nanostrip for 5 minutes, then rinsed 3 times in DI water. The chips are placed in a vacuum desiccator. 5 drops of trichloro(1H,1H,2H,2H-perfluorooctyl)silane (PFOTS, Sigma Aldrich) are placed in a cap in the bottom of the desiccator. The desiccator is evacuated slowly, over the course of 3 minutes, then isolated from the vacuum pump and left evacuated for 1 hour. The chips are removed from the vacuum desiccator. 20 nm of Au is evaporated in an electron-beam evaporator at a speed of 1 Å/s and a base pressure below 5E-7 Torr. Photolithography is performed using positive photoresist and a positive photomask to define the contacts. For photolithography, S1813 is used according to the following recipe: spin at 4000 rpm for 30 seconds, soft bake at 115ºC for 1 minute, expose to 365 nm UV light at 15 mW/cm² for 8 seconds, develop in MF 319 for 50 seconds, and then rinse in DI water for 10 seconds. The Au around the photoresist is etched by immersion of the chips in Transene Gold Etchant TFA for 10 seconds, then the sample is rinsed three times in DI water. The photoresist is dissolved in slightly heated acetone (60ºC for 5 minutes).

*Step 2: PDMS/PPC stamps.* A similar procedure is followed to that developed by Pizzochero *et al*[23]. PDMS (Sylgard 184) is mixed in a glass petri dish and left in an oven at 60-70ºC overnight. PPC is made by stirring 1.5 g PPC in 10 mL anisole on a hot plate at 60ºC for 1 hour. The PDMS is cut into 1 cm by 1 cm squares with a razor blade. One square is removed from the petri dish, rinsed in IPA for 20 seconds, then dried in nitrogen gas. The PDMS stamp is placed on one end of a glass slide, and the thickest corner of the stamp is identified. The PDMS stamps are plasma-ashed at 300 mTorr and 120 W for 10 minutes. The PDMS stamp is centered on the spinner, and 2 drops of PPC are placed on the PDMS stamp, and then spun at 1500 rpm for 1 minute. The PDMS/PPC stamps are let sit for a few minutes, but not longer than 10 minutes. The edges of the PDMS/PPC stamp are cut away with a fresh razor blade until the stamp is ~2 mm by 2 mm.

*Step 3: Transfer of metal contacts.* The stage of a 2D transfer setup is heated to 60ºC. A Si/SiO$_2$ chip containing metal contacts is loaded onto the stage, and the desired contact is centered in the field of view. The PDMS/PPC stamp is loaded onto the top arm of the transfer setup. The thickest corner of the PDMS/PPC stamp is centered in the field of view. When the stamp is lowered, the polymer front should originate from this corner. The PDMS/PPC stamp and the desired contact are aligned so that the PPC completely covers the contact, but so that just the corner of the stamp will make contact with the substrate. The PDMS/PPC stamp is lowered slowly. Once the polymer front progresses just past the contact, the stage temperature is lowered to 40ºC. After the temperature has reached 40ºC, the top arm of the transfer setup is raised slowly. As the polymer front begins to move, but before it reaches the contact, the top arm of the transfer setup and therefore the PDMS/PPC stamp is raised very quickly, picking up the contact with it. The stage of the transfer setup is heated to 60ºC again. The target substrate is loaded onto the stage of the transfer setup and the target flake is centered in the field of view. Once the temperature reaches 60ºC, the contact on the PDMS/PPC stamp is aligned with the flake. The PMDS/PPC stamp is lowered slowly, until the polymer front progresses just past the sample. Immediately, and with the stage still at 60ºC, the PDMS/PPC stamp is raised very slowly. The contact should delaminate from the stamp and stick to the flake. PPC that occasionally sticks to the flake can be removed by rinsing in chloroform for 5 minutes, then drying in nitrogen gas.

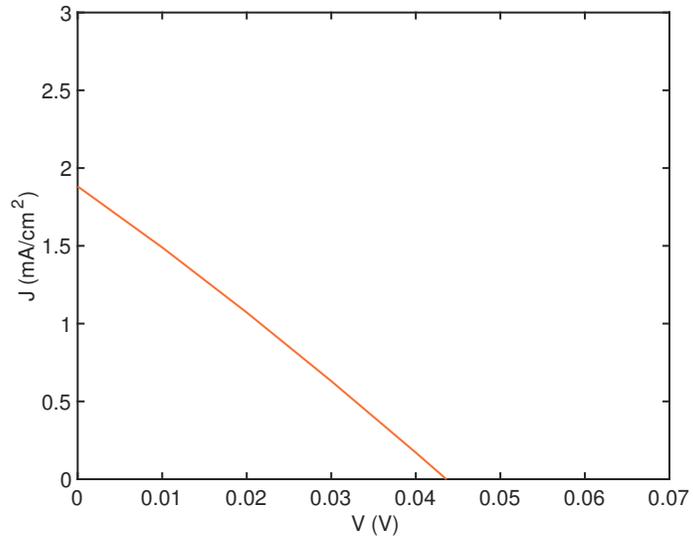

**Supplementary Fig. 1** Simulated I-V curve for evaporated devices. Simulations can replicate the resistive behavior of evaporated-contact devices, assuming the metal contacts have an effective work function difference of 50 meV due to Fermi-level pinning at the Au contact.

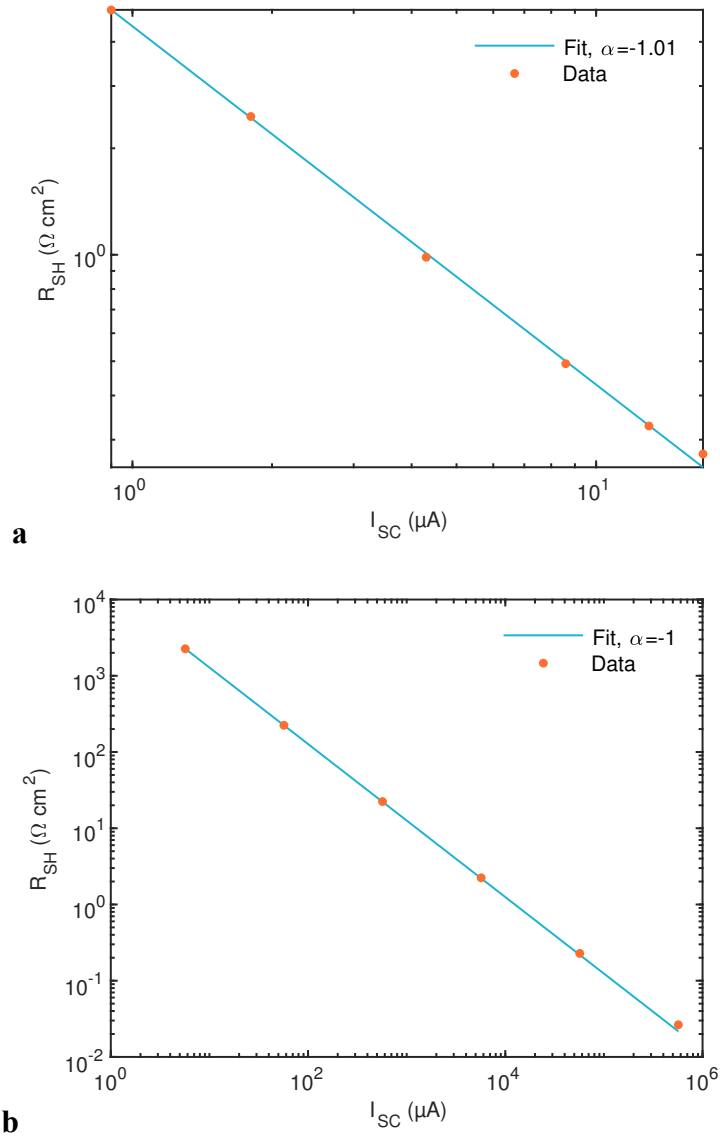

**Supplementary Fig. 2** Photoshunting. **a** $R_{SH}$ vs. $I_{SC}$, experimental. $R_{SH}$ is extracted from the slopes of the power-dependent I-V curves (Fig. 3a of the main text) at short-circuit. A power law fit yields $\alpha \approx -1$. **b** $R_{SH}$ vs. $I_{SC}$, simulated. A power law fit yields $\alpha = -1$, suggesting that the shunt behavior is well-explained by increasing minority carrier conductivity at higher laser powers and higher short-circuit currents.

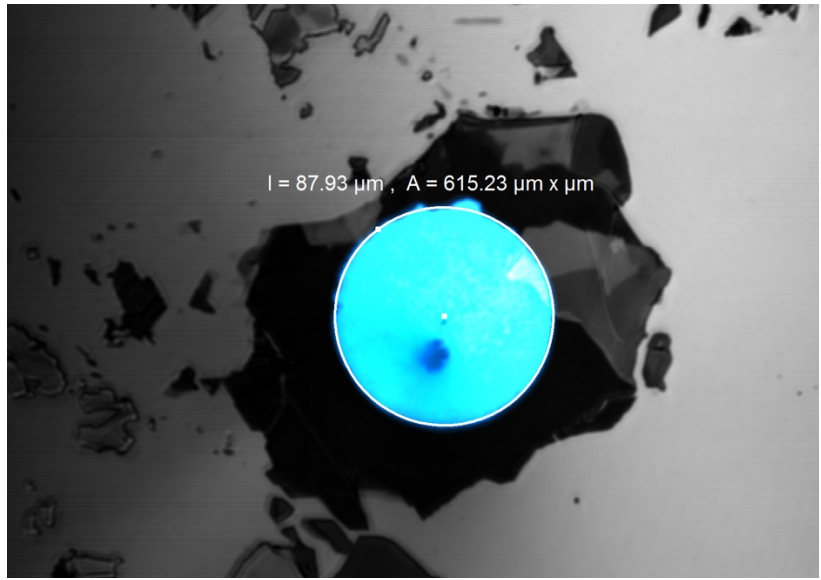

**Supplementary Fig. 3** Active area. The active area of the device presented in the main text is measured to be 615 μm².

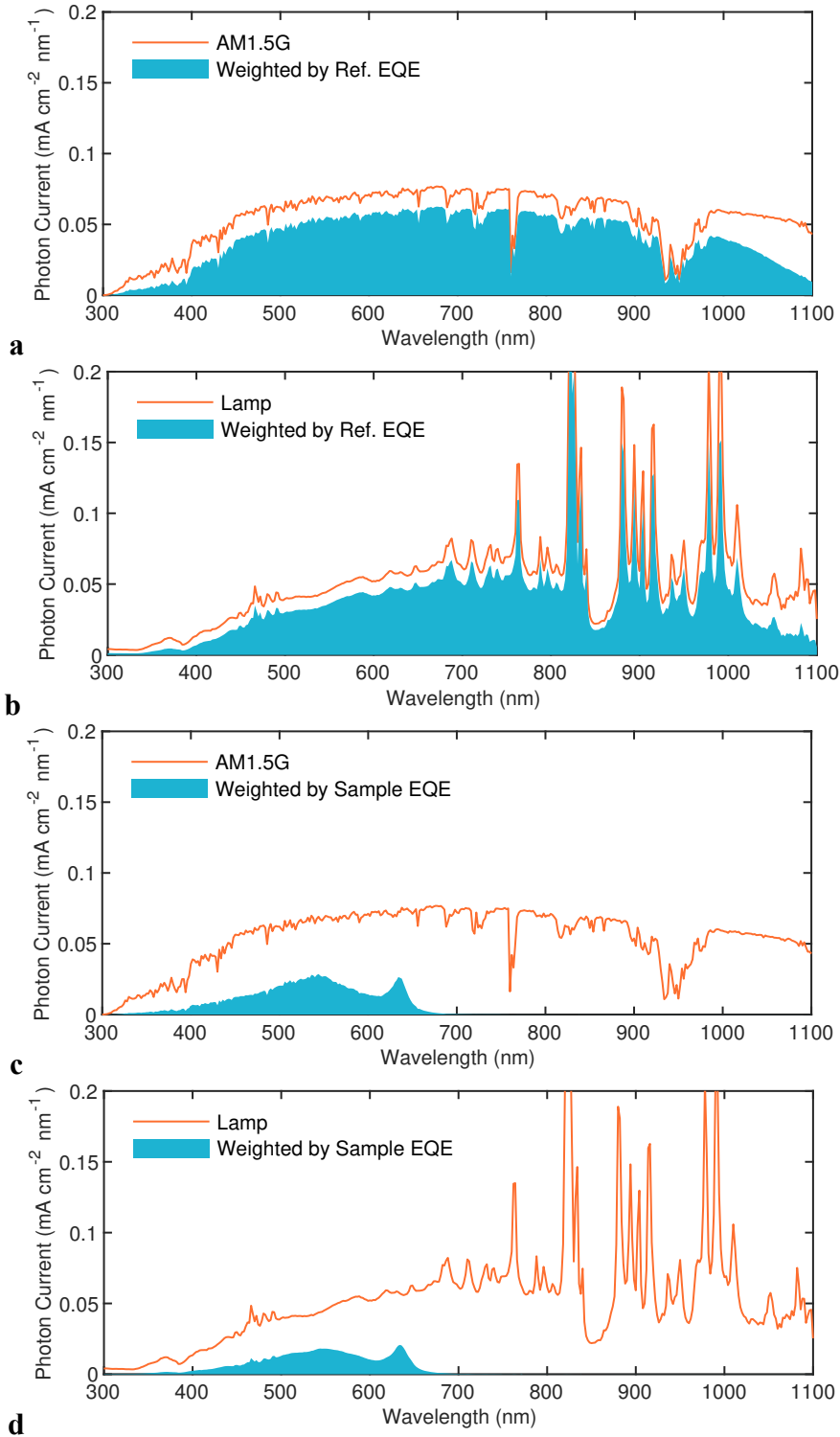

**Supplementary Fig. 4** Spectral mismatch. **a, b** Current generated under AM1.5G (**a**) and solar simulator lamp (**b**) for our silicon reference solar cell. We adjust the lamp intensity until these integrated currents are equal. **c, d** Same, but for the sample. We calculate the spectral mismatch factor M = 0.67 by dividing the integrated current in (**d**) by the integrated current in (**c**).

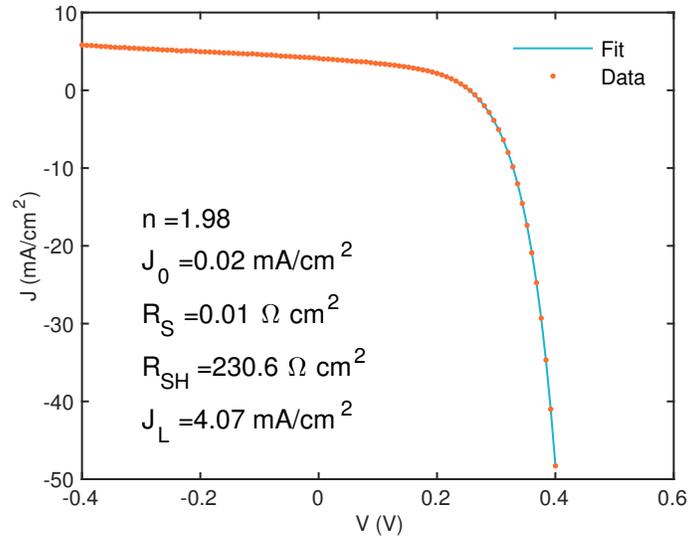

**Supplementary Fig. 5** Fitting for one-sun I-V curve. The diode equation with shunt and series resistances is used to extract the parasitic resistances and diode parameters.

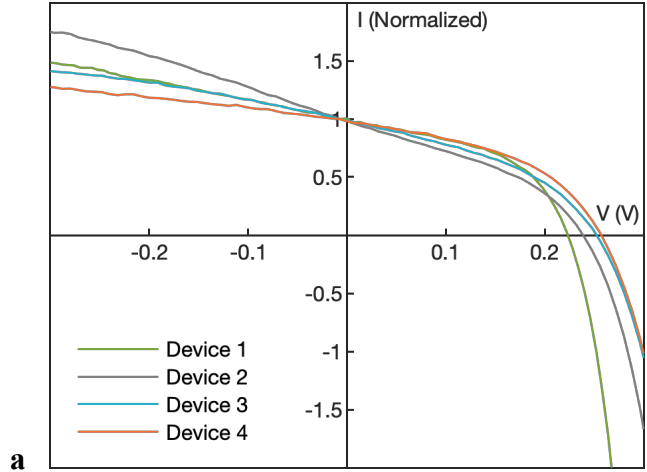

a

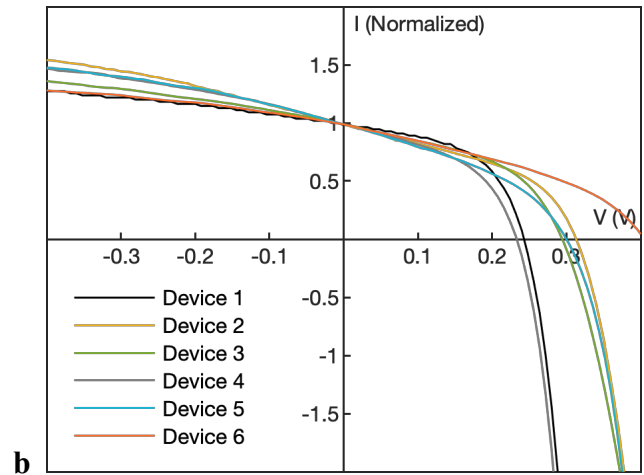

b

**Supplementary Fig. 6** Replicability. **a** I-V curves taken from 4 different devices under one-sun illumination. $V_{OC}$ is between 220 and 250 mV across all devices studied. **b** I-V curves taken from 6 different devices under illumination with a halogen lamp (~20 suns power density). $V_{OC}$ is greater than 220 mV across all devices studied. In both one-sun and halogen lamp I-V curves, $I_{SC}$ varies due to the different thicknesses and active areas across different devices, and therefore current is normalized by the short-circuit current value for easier comparison.

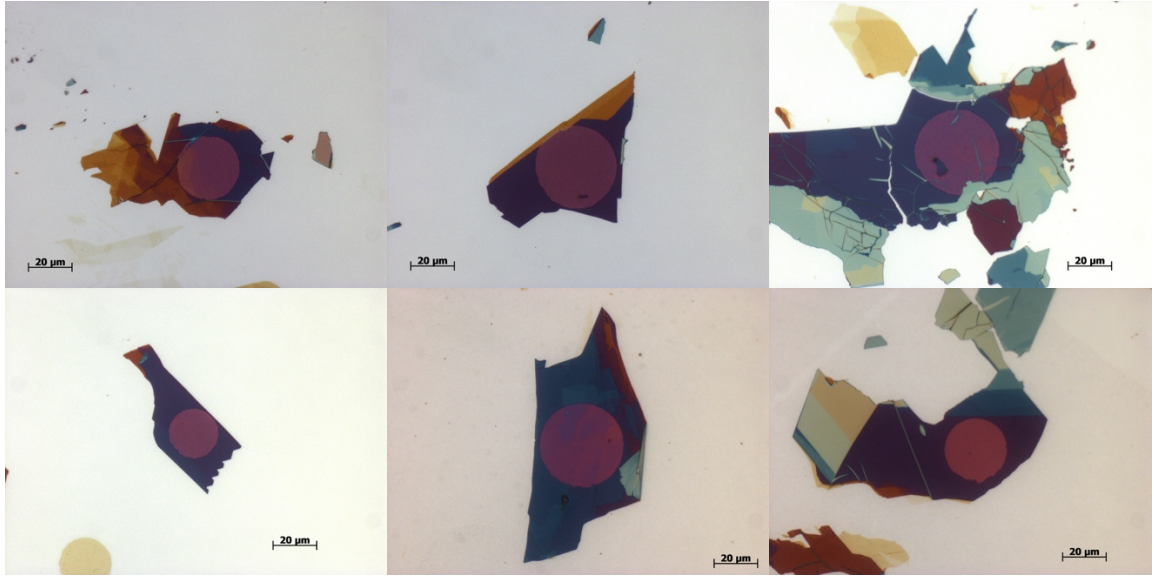

**Supplementary Fig. 7** Microscope images of other fabricated devices. The first 5 images are Devices 1-5 in Supplementary Fig. 5b.

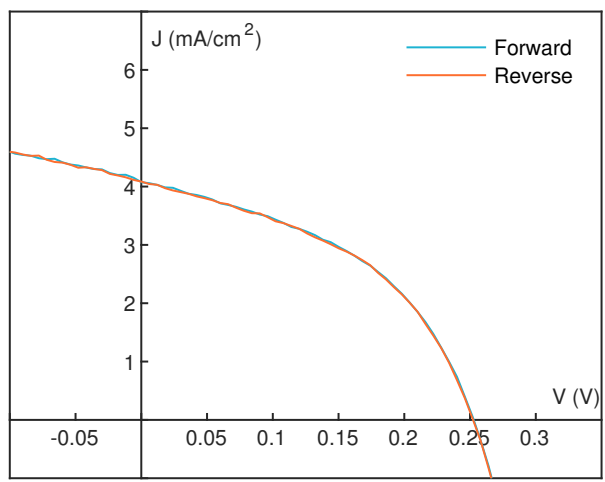

**Supplementary Fig. 8** Forwards/backwards scans. One-sun I-V curves swept in forward and reverse directions show no hysteresis.

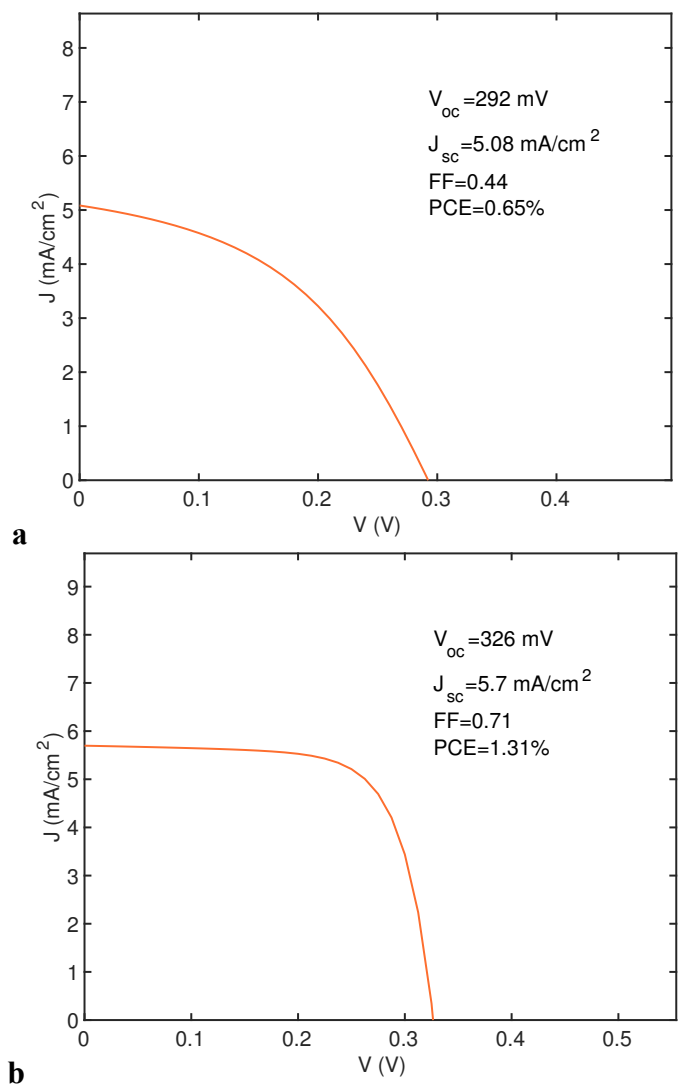

**Supplementary Fig. 9** Matching simulations to experimental device. **a** An I-V curve with parameters close to the experimentally-observed I-V curve can be simulated if the Au work function is set to 4.6 eV, while all other parameters are kept the same. $R_{SH}$ for this simulation, approximated by the inverse slope of the I-V curve at short-circuit, is 275 $\Omega$ cm². Au could have a lower effective work function due to contamination or thiol bonding at the Au/WS$_2$ interface. However, this would also limit the $V_{OC}$ observed under laser illumination to ~400 mV and thus is not a complete explanation. **b** Increasing mobility reduces the $V_{OC}$, but does not change the shunt resistance. Here, mobility is set to 1000 cm² V⁻¹ s⁻¹ for both electrons and holes. $R_{SH}$ for this simulation, approximated by the inverse slope of the I-V curve at short-circuit, is 2238 $\Omega$ cm².

**Supplementary Table 1: WS$_2$ Parameters for Device Simulations**

| Parameter | Value | Source |
|---|---|---|
| Bandgap | 1.35 eV | Ref. 30 |
| Work function | 4.62 eV | Ref. 31 |
| Electron effective mass | 0.63m$_e$ | Ref. 33 |
| Hole effective mass | 0.84m$_e$ | Ref. 33 |
| Doping | 10$^{14}$ cm$^{-3}$ | HQ Graphene |
| Out-of-plane mobility (for holes & electrons) | 0.01 cm$^2$/Vs | Ref. 34–36 |
| DC permittivity | 6.7 | Ref. 32 |
| Radiative recombination coefficient | 1.64×10$^{-14}$ cm$^3$/s | Calculated using optical constants in Ref. 29 |
| Shockley-Read-Hall Lifetime | 6.17 ns | Calculated using PLQY from Ref. 37–39 |